\documentclass[12pt,preprint]{aastex}

\begin{document}
\slugcomment{ApJ in press; submitted 2006 September 22}
\title{Turbulent Mixing and the Dead Zone in Protostellar Disks}

\author{N.~J.~Turner} \affil{Jet Propulsion Laboratory MS 169-506,
  California Institute of Technology, Pasadena, California 91109, USA;
  neal.turner@jpl.nasa.gov}
\author{T.~Sano} \affil{Institute of Laser Engineering, Osaka
  University, Suita, Osaka 565-0871, Japan; sano@ile.osaka-u.ac.jp}
\and
\author{N.~Dziourkevitch} \affil{Max Planck Institute for Astronomy,
  K\"{o}nigstuhl 17, D-69117 Heidelberg, Germany; natalia@mpia-hd.mpg.de}

\begin{abstract}
  We investigate the conditions for the presence of a magnetically
  inactive dead zone in protostellar disks, using 3-D shearing-box MHD
  calculations including vertical stratification, Ohmic resistivity
  and time-dependent ionization chemistry.  Activity driven by the
  magnetorotational instability fills the whole thickness of the disk
  at 5~AU, provided cosmic ray ionization is present, small grains are
  absent and the gas-phase metal abundance is sufficiently high.  At
  1~AU the larger column density of 1700~g~cm$^{-2}$ means the
  midplane is shielded from ionizing particles and remains
  magnetorotationally stable even under the most favorable conditions
  considered.  Nevertheless the dead zone is effectively eliminated.
  Turbulence mixes free charges into the interior as they recombine,
  leading to a slight coupling of the midplane gas to the magnetic
  fields.  Weak, large-scale radial fields diffuse to the midplane
  where they are sheared out to produce stronger azimuthal fields.
  The resulting midplane accretion stresses are just a few times less
  than in the surface layers on average.
\end{abstract}


%%%%%%%%%%%%%%%%%%%%%%%%%%%%%%%%%%%%%%%%%%%%%%%%%%%%%%%%%%%%%%%%%%%%%%%%%%%%%%%
\section{INTRODUCTION\label{sec:intro}}

Turbulence plays a central role in protostellar disks.  It transports
orbital angular momentum, controlling the accretion of gas on the
central star \citep{ss73}; brings dust grains together in the early
stages of planet formation \citep{wc93}; and modifies the orbital
migration of protoplanets \citep{np04}.  Turbulence can be driven by
the magneto-rotational instability (MRI; Balbus \& Hawley 1991)
%\cite{bh91}
in regions where the ionization is sufficient to couple the gas to the
magnetic fields.  The MRI operates by transferring orbital angular
momentum outward along magnetic field lines linking gas on different
orbits.  The fastest linear growth rate is three-quarters the orbital
frequency $\Omega$ and occurs at wavelengths near $\lambda_c = 2\pi
v_A/\Omega$ in fully-ionized Keplerian disks with Alfven speed $v_A$
\citep{bh91}.  In the non-linear regime, the gravitational energy
released from the inspiraling gas sustains turbulence and regenerates
the magnetic fields \citep{bh98}.  Good coupling of the fields to the
gas within a few tenths of an AU of the star comes from thermal
ionization of the low-ionization-potential elements potassium and
sodium \citep{pm65,u83}.  At greater distances, the ionization is due
to the absorption of stellar X-rays or interstellar cosmic rays in the
disk material.  The X-rays penetrate to columns of about 10
g~cm$^{-2}$ \citep{ig99} and the cosmic rays to 100 g~cm$^{-2}$, while
recombination is generally fastest in the dense interior, leading to a
layered structure.  The top and bottom surfaces of the disk are
magnetically active.  The interior is quiescent and called the dead
zone \citep{g96}.  The dead zone is of special interest because it
overlaps the region where the planets are found and potentially
provides the environment and the raw ingredients for planet formation.
The simplest layered picture, however, yields an accretion rate
independent of stellar mass, conflicting with data for T~Tauri stars.
Including the extra heating from irradiation of the disk by the star
gives a mass dependence that is still too weak \citep{hd06},
indicating that our understanding of the dead zone is incomplete.

The size of the dead zone depends on the disk surface density, the
abundance of the dust that hastens recombination \citep{wn99,sm00},
the presence of long-lived metal ions \citep{ft02} and the rate at
which the turbulence mixes the charged species into the interior
\citep{is05,in06b}.  Well-mixed 0.1-$\mu$m grains with the
interstellar dust-to-gas mass ratio in the minimum-mass Solar nebula
produce a dead zone extending to about 20~AU \citep{sm00,mp06}.  Even
if small grains are absent, the annulus immediately outside the
thermally-ionized region is quiescent for a wide range of disk models
because the high surface densities mean cosmic rays and X-rays are
absorbed before reaching the midplane.  In this paper we investigate
whether the dead zone can be eradicated under favorable conditions
leaving the entire disk active.  Because cosmic rays penetrate more
deeply, their ionizing effects are included despite possible screening
by the protostellar wind.  Slow recombination is ensured by assuming
the dust surface area per unit gas mass has declined through settling,
coagulation and orbital evolution sufficiently that recombination on
grain surfaces can be neglected.  Since the abundance of gas-phase
metal atoms is not well known, it is treated as a parameter.  We focus
on conditions near the present orbits of the Earth and Jupiter and use
the minimum-mass Solar nebula \citep{hn85} as the disk model.  We
extend results from \cite{fs03} by making stratified shearing-box MHD
calculations including time-dependent ionization chemistry.

The paper has six sections.  The ionization-recombination reaction
network, disk model and MHD methods are described in
section~\ref{sec:methods}.  The network is used in
section~\ref{sec:timescales} to estimate the size of the dead zone in
local chemical equilibrium.  Expansion or contraction of the dead zone
through turbulent mixing is explored using MHD calculations in
section~\ref{sec:mixing} while section~\ref{sec:criteria} deals with
simple expressions for predicting the boundaries of the dead zone.  A
summary and conclusions are in section~\ref{sec:conc}.

%%%%%%%%%%%%%%%%%%%%%%%%%%%%%%%%%%%%%%%%%%%%%%%%%%%%%%%%%%%%%%%%%%%%%%%%%%%%%%%
\section{METHODS\label{sec:methods}}

\subsection{Reaction Network\label{sec:reactions}}

The ionization state of the disk gas is calculated using the
simplified reaction network developed for molecular cloud studies by
\cite{od74}.  Five species are involved: molecular hydrogen and its
ion; magnesium, a representative metal, and its ion; and free
electrons.  These take part in the four reactions
\begin{eqnarray}
{\rm H}_2            &\rightarrow& {\rm H}_2^++e^-\label{eqn:ionization}\\
{\rm H}_2^++e^-      &\rightarrow& {\rm H}_2\label{eqn:dissrec}\\
{\rm H}_2^++{\rm Mg} &\rightarrow& {\rm H}_2+{\rm Mg}^+\label{eqn:chgexch}\\
{\rm Mg}^+ +e^-      &\rightarrow& {\rm Mg}+h\nu.\label{eqn:radrec}
\end{eqnarray}
The ionization (eq.~\ref{eqn:ionization}) is balanced by both
dissociative recombination of the molecular ion
(eq.~\ref{eqn:dissrec}) and molecule-metal charge exchange
(eq.~\ref{eqn:chgexch}) followed by the radiative recombination of
the metal ion (eq.~\ref{eqn:radrec}).  The rate coefficients are
$3\times 10^{-6 }/\sqrt{T}$ cm$^3$~s$^{-1}$ for dissociative recombination, 
$3\times 10^{-9 }         $ cm$^3$~s$^{-1}$ for charge exchange and
$3\times 10^{-11}/\sqrt{T}$ cm$^3$~s$^{-1}$ for radiative recombination
at a temperature $T$ in Kelvin \citep{in06a}.  Because radiative
recombination is slowest, the positive charge typically resides on
metal ions in chemical equilibrium.  Comparison with a comprehensive
network of almost 2000~reactions shows that the Oppenheimer \&
Dalgarno network yields slightly higher equilibrium electron
abundances because fewer recombination pathways are included
\citep{in06a}.  A similar reduced reaction set adequately reproduces
the ionization fraction from a detailed chemical model of the midplane
of a protostellar disk at 1~AU \citep{sw04}.  The kinetic equations
for the reactions~\ref{eqn:ionization}-\ref{eqn:radrec} form a stiff
set of ODEs that is integrated using semi-implicit extrapolation
\citep{pt92}.  Solving the reaction network uses about ten times as
many CPU cycles per zone as evolving the magnetic fields.

The cosmic ray ionization rate is reduced from the interstellar value
$\zeta_{CR}=10^{-17}$ s$^{-1}$ \citep{cw98} by the shielding effects
of the surrounding disk material.  The total ionization rate
\begin{equation}
\zeta(z) = {\zeta_{CR}\over 2} \left\{
\exp\left[-\chi^+(z)\over\chi_{CR}\right] +
\exp\left[-\chi^-(z)\over\chi_{CR}\right] \right\}
\end{equation}
depends on the cosmic ray absorption depth $\chi_{CR}=96$ g~cm$^{-2}$
and the column above and below the point of interest, $\chi^+ =
\int_z^{+\infty} \rho\,dz$ and $\chi^- = \int_{-\infty}^z \rho\,dz$.
In the 3-D MHD calculations of section~\ref{sec:mixing}, the column
densities are averaged over the horizontal extent of the domain when
finding the ionization rate.

The resistivity is computed from the electron abundance by
\begin{equation}\label{eqn:resistivity}
\eta = 234\,{\sqrt{T}\over x_e}\,{\rm cm^2 s^{-1}},
\end{equation}
where $x_e=n_e/n_n$ is the electron fraction, $n_e$ the electron
number density and $n_n$ the total number density of neutrals
\citep{bb94}.

\subsection{Disk Model\label{sec:disk}}

Our model disk is the minimum-mass Solar nebula in which the surface
density and temperature vary with radius $r$ as $\Sigma = 1700 (r/{\rm
  AU})^{-3/2}$ g~cm$^{-2}$ and $T = 280 (r/{\rm AU})^{-1/2}$~K,
respectively.  The temperature is independent of the distance from the
midplane, so that in hydrostatic equilibrium the density falls off
with height $z$ as $\exp(-z^2/2H^2)$.  The scale height $H=c_s/\Omega$
depends on the sound speed $c_s=({\cal R}_{gas}T/\mu)^{1/2}$.

The composition of the material is Solar \citep{ag05}, with
0.085~helium and $3.39\times 10^{-5}$ magnesium atoms per hydrogen
nucleus.  The hydrogen is molecular and the mean molar weight
$\mu=2.29$.  Only gas-phase magnesium contributes to the overall
ionization level, while the magnesium atoms can take three forms, (1)
incorporated into silicate and metallic grains, (2) adsorbed on grain
surfaces and (3) in the gas phase.  Measurements of the diffuse
interstellar gas show that about half the magnesium is in grains
\citep{sc94,f97}.  We assume the remainder is divided between grain
surfaces and the gas phase through thermal adsorption and desorption
and we calculate the equilibrium gas-phase fraction as outlined by
\cite{in06a}.  The equilibrium is reached quickly, as the lifetime of
adsorbed magnesium atoms is just $10^{-4}$~seconds at 1~AU and 280~K,
or $10^6$~seconds at 5~AU and 125~K.  Almost all the magnesium not in
grain interiors is in the gas phase at 1~AU, while at 5~AU just 1.4\%
stays in the gas if 0.1-micron grains are present with the
interstellar mass fraction.  If the gas-phase magnesium abundance is
unchanged by the removal of the small grains, its most probable value
is half Solar at 1~AU and 0.7\% Solar at 5~AU.  We consider a range of
possible magnesium abundances below.

The evolution of the magnetic fields is potentially affected by the
Ohmic, Hall and ambipolar diffusion terms in the induction equation.
We estimate the magnitudes of the terms as outlined in \cite{ss02a}.
The Ohmic and Hall terms are comparable near the midplane, while the
ambipolar term is the largest of the three only above about $3H$.
Linear analyses \citep{w99,bt01,sw05} indicate the Hall term has
little effect on the maximum MRI growth rate but extends the range of
unstable wavelengths.  Numerical results suggest the strength of the
saturated turbulence is largely unaffected by the Hall term and is
determined by the Ohmic term \citep{ss02b}.  The ambipolar term
decouples the neutral gas from the ionized component \citep{hs98} at
the low densities found in our model disk outside $|z|=4H$, where
collisions transfer momentum from the ions to the neutrals at rates
less than the orbital frequency.  Little accretion of neutral gas due
to MRI turbulence will occur beyond four scale heights.  Since the
Ohmic term is most important for the size of the dead zone, we choose
to focus on the effects of the Ohmic resistivity.

\subsection{MHD Equations\label{sec:mhd}}
 
The standard equations of resistive MHD,
\begin{equation}\label{eqn:cty}
{{\partial\rho}\over{\partial t}}+{\bf\nabla\cdot}(\rho{\bf v})=0,
\end{equation}
\begin{equation}\label{eqn:eomg}
{{\partial{\bf v}}\over{\partial t}} + {\bf v \cdot \nabla v} =
	- {{\bf\nabla}p\over\rho}
	+ {1\over 4\pi\rho}({\bf\nabla\times B}){\bf\times B}
	- 2{\bf\Omega\times v} + 3\Omega^2 {\bf\hat x}
	- \Omega^2 {\bf\hat z},
\end{equation}
and
\begin{equation}\label{eqn:dbdt}
{\partial{\bf B}\over\partial t} = {\bf\nabla\times}({\bf v\times B}
  - \eta{\bf\nabla\times B}),
\end{equation}
are solved with an isothermal equation of state $p=\rho c_s^2$, using
the {\sc ZEUS} code \citep{sn92a,sn92b}.  When the chemical reactions
are included, an additional continuity equation
\begin{equation}\label{eqn:ctyi}
{{\partial\rho_i}\over{\partial t}}+{\bf\nabla\cdot}(\rho_i{\bf v})=0,
\end{equation}
is solved for each of the five chemical species $i$.  The domain is a
three-dimensional shearing-box \citep{hg95} and vertical
stratification is included.  The radial or $x$-boundaries are
periodic, with an azimuthal offset that increases in time according to
the radial gradient in orbital speed.  The azimuthal or $y$-boundaries
are strictly periodic, while the vertical or $z$-boundaries allow
outflow but no inflow.  The domain extends two density scale heights
$H$ along the radial direction, eight along the orbital direction and
four either side of the midplane and is divided into $32\times
64\times 128$ grid zones.  Short timesteps are avoided by applying a
density floor of $1/15$ the minimum initial density.  The floor
typically operates in just a few grid zones near domain top and
bottom, and for all the calculations described in this paper the mass
added is much less than the mass escaping through the boundaries,
which in turn is in most cases less than 1\% of the total initial mass
per 30~orbits.  The starting magnetic field has strength $B_0$ chosen
so that the midplane MRI wavelength is $0.5 H$.  The components of the
field are $(B_x, B_y, B_z) = [0, B_0\sin(\pi x/H), B_0\cos(\pi x/H)]$.
Magnetic pressure and tension forces initially vanish, as the field
strength is uniform and the field lines are straight.  The net
magnetic flux is zero because the field lines rotate $360^\circ$ round
the $y-z$ plane across the radial extent of the box.  Due to the
periodic conditions on the side boundaries, the vertical component of
the magnetic flux is conserved during the calculations.  The outflow
conditions on the top and bottom boundaries mean that, unlike the
calculations discussed by \cite{fs03}, the radial and azimuthal
magnetic fluxes are free to evolve over time.

%%%%%%%%%%%%%%%%%%%%%%%%%%%%%%%%%%%%%%%%%%%%%%%%%%%%%%%%%%%%%%%%%%%%%%%%%%%%%%%
\section{TIMESCALES\label{sec:timescales}}

To gain an overall grasp of the problem, we estimate three timescales
affecting the size of the dead zone.  These are (1) the Ohmic
dissipation time for the magnetic fields, (2) the turbulent mixing
time and (3) the time for ionized gas to recombine and reach chemical
equilibrium.  The dead zone is the region where the magnetic fields
dissipate too fast for the MRI to grow.  The dissipation time is
calculated in this section assuming local chemical equilibrium so that
the electron abundance at each height results solely from local
ionization and recombination processes.  However the dead zone
boundaries will shift if ionized gas from the surface layers is mixed
to the interior before recombining.  The three corresponding
timescales are calculated as follows.
\begin{enumerate}
\item Like any diffusion timescale, the dissipation and mixing times
  are ratios of a squared length to a diffusion coefficient.  The
  diffusion coefficient $\eta$ for resistive dissipation is given by
  eq.~\ref{eqn:resistivity}.  For the length we choose the scale
  height $H$, yielding an upper bound on the dissipation time in the
  disk interior, where the magnetic fields can alternate in sign over
  distances shorter than the scale height.  MRI turbulence is expected
  when the dissipation time is longer than the growth time of the
  instability.  For example if the characteristic unstable wavelength
  $\lambda_c$ is one-tenth of the scale height then turbulence
  requires a dissipation time, $H^2/\eta$, longer than about
  $(0.1)^{-2} = 100$~orbits.  The diffusion time in orbits can be
  converted to a magnetic Reynolds number $c_s^2/(\eta\Omega)$ by
  multiplying by $2\pi$.
\item The mixing time depends on the turbulent diffusion coefficient,
  which is the product of the mean squared velocity fluctuation and
  the eddy correlation time \citep{fp06}.  For a preliminary estimate
  of the mixing time, these quantities are taken from ideal-MHD
  shearing-box calculations, with the Ohmic resistivity switched off
  in eq.~\ref{eqn:dbdt}.
\item The recombination time is measured in the hydrostatic model disk
  starting with the magnesium and electron abundances equal, the
  magnesium fully-ionized and the molecular hydrogen neutral.  The
  reaction network is then integrated until the electron abundance
  reaches equilibrium.
\end{enumerate}
We consider conditions at 1 and 5~AU in the minimum-mass Solar nebula.
The mixing times are found using turbulent diffusion coefficients from
the ideal-MHD calculations M1 and M5 listed in table~\ref{tab:parms}.
The coefficients are averaged over the horizontal extent of the domain
and over a period from 10 to 100~orbits after the turbulence is
well-established.
%\clearpage
\begin{deluxetable}{rlrrr}
\tablewidth{0pt}
\tablecaption{Calculation Parameters\label{tab:parms}}
\tablehead{Label & Physics & $R/{\rm AU}$ & $x_{\rm Mg}/{\rm Solar}$
 & Duration/orbits}
\startdata
M1  & Ideal MHD               & 1 & --        & 100 \\
F1  & Fixed $\eta(z)$         & 1 & 1         & 150 \\
V1  & Varying $\eta(x,y,z,t)$ & 1 & 1         & 150 \\
\\
M5  & Ideal MHD               & 5 & --        & 100 \\
F52 & Fixed $\eta(z)$         & 5 & $10^{-2}$ & 100 \\
F56 & Fixed $\eta(z)$         & 5 & $10^{-6}$ & 100 \\
F58 & Fixed $\eta(z)$         & 5 & $10^{-8}$ & 100 \\
\enddata
\end{deluxetable}
%\clearpage
\subsection{Timescales at 1~AU\label{sec:timescales1au}}

\begin{figure}[tb!]
  \epsscale{0.7}
  \plotone{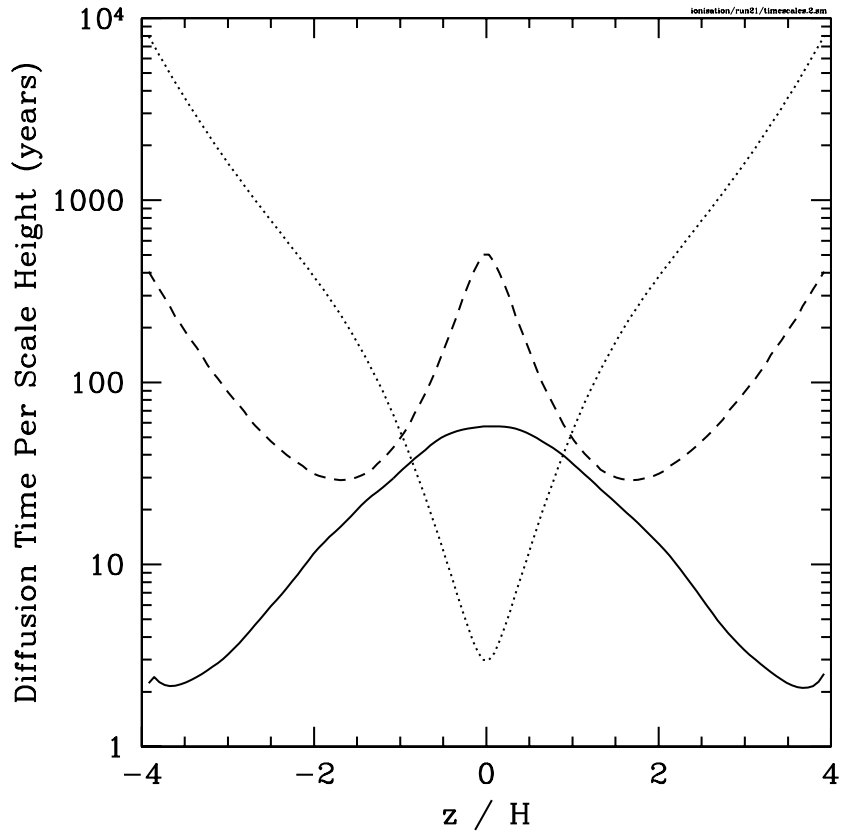}
  \figcaption{Timescales for Ohmic dissipation (dotted), recombination
    (dashed) and turbulent mixing (solid) as functions of height at
    1~AU, assuming Solar abundance for magnesium in the gas phase.
    The mixing timescale is taken from the ideal-MHD calculation M1.
    \label{fig:z1au}}
\end{figure}

\begin{figure}
  \epsscale{0.7}
  \plotone{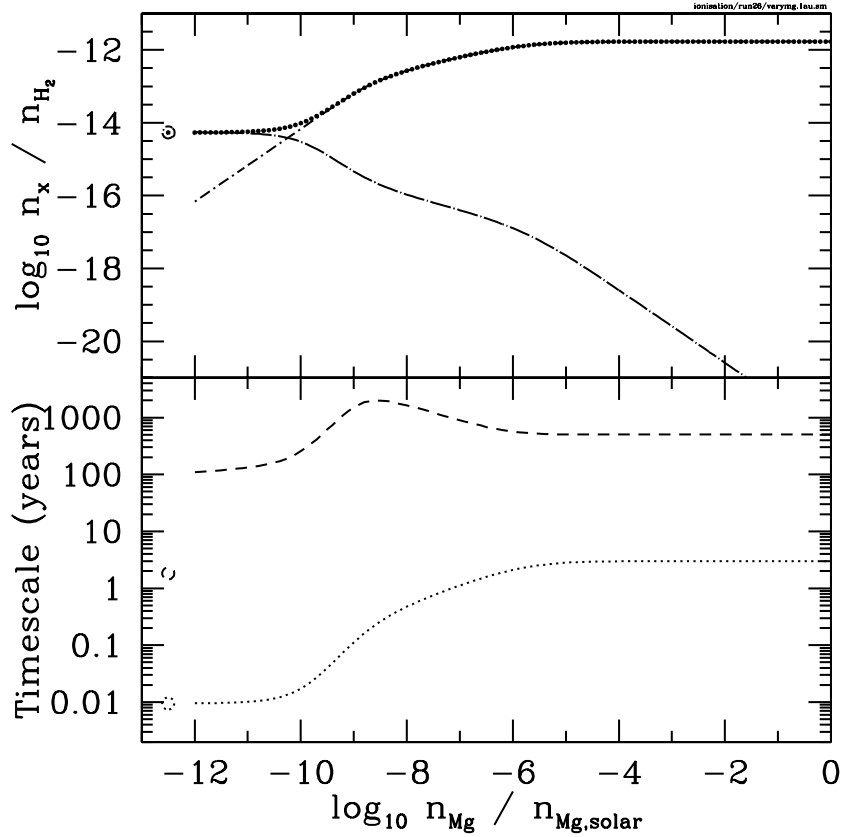}
  \figcaption{Midplane charged particle abundances (top panel) and
    timescales (bottom) versus gas-phase magnesium abundance at 1~AU.
    Above, H$_2^+$ is shown by a long-dash-dotted line, Mg$^+$ by a
    short-dash-dotted line and electrons by small circles.  Below, the
    recombination timescale is shown by a dashed line and the Ohmic
    dissipation time per scale height by a dotted line.  Circles near
    the left-hand edge mark the values when magnesium is absent.
  \label{fig:mg1au}}
\end{figure}

Results for 1~AU are shown in figure~\ref{fig:z1au}.  Because we aim
to find the size of the active region under the most favorable
conditions, we assume here that the gas-phase magnesium abundance is
Solar.  Nevertheless the magnetic fields dissipate in just three
orbits at the midplane if the ionization is in equilibrium.  The MRI
is absent near $z=0$ as all modes fitting in the disk thickness
\citep{sm99} are damped.  The recombination time is several hundred
years at the midplane and also at four scale heights, but is only
thirty years at two scale heights, where the absolute number density
of free electrons is greatest.  At all heights, the turbulent mixing
from the ideal-MHD calculation is faster than the recombination,
suggesting mixing could affect the size of the dead zone.
%\clearpage
%\clearpage
The recombination and dissipation times depend on the metal abundance
as shown in figure~\ref{fig:mg1au}.  The midplane Ohmic dissipation
time is 3~orbits or less across the entire range of abundances
considered, indicating that changes in the metal abundance alone will
not remove the dead zone.  In contrast, at $z=2H$ the dissipation time
is longer than 100~orbits and magnetic activity is expected if the
magnesium abundance is greater than $10^{-5}$ Solar.  The dissipation
and recombination times at all heights are abundance-independent above
$10^{-4}$ Solar, so the results with Solar magnesium accurately
reflect conditions throughout the likely abundance range: in local
chemical equilibrium, the surface layers are magnetically active while
the midplane is dead.

%\clearpage
\subsection{Timescales at 5~AU\label{sec:timescales5au}}

\begin{figure}[tb!]
  \epsscale{0.7}
  \plotone{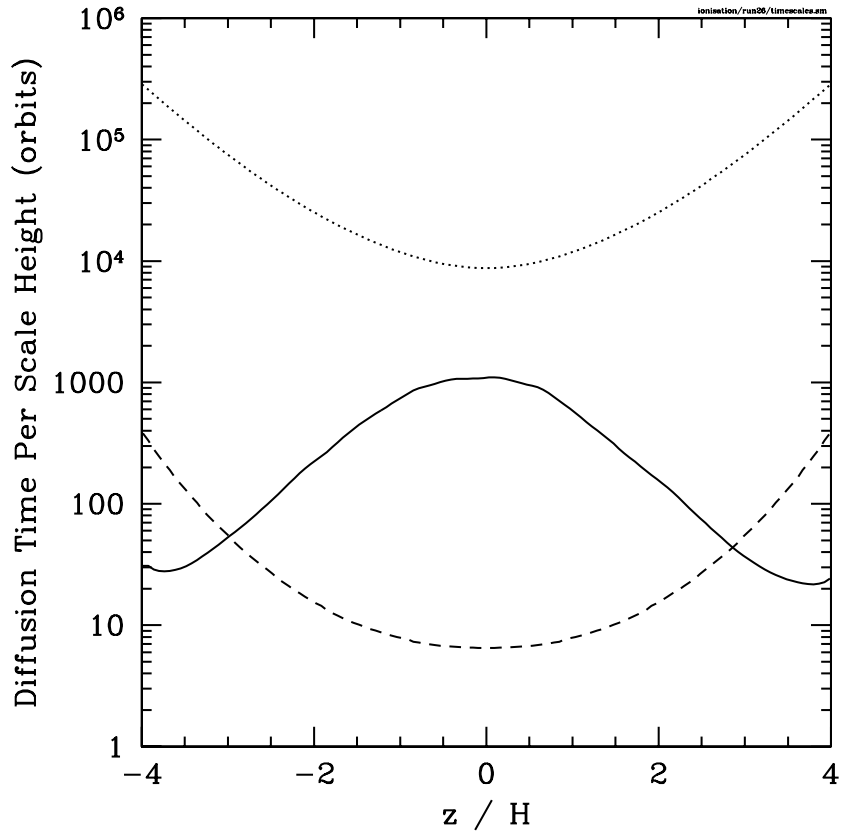}
  \figcaption{Timescales for Ohmic dissipation (dotted), recombination
    (dashed) and turbulent mixing (solid), as functions of height at
    5~AU, with a gas phase magnesium abundance $10^{-2}$ Solar.  The
    mixing timescale is taken from the ideal-MHD calculation M5.
    \label{fig:z5au}}
\end{figure}

\begin{figure}
  \epsscale{0.7}
  \plotone{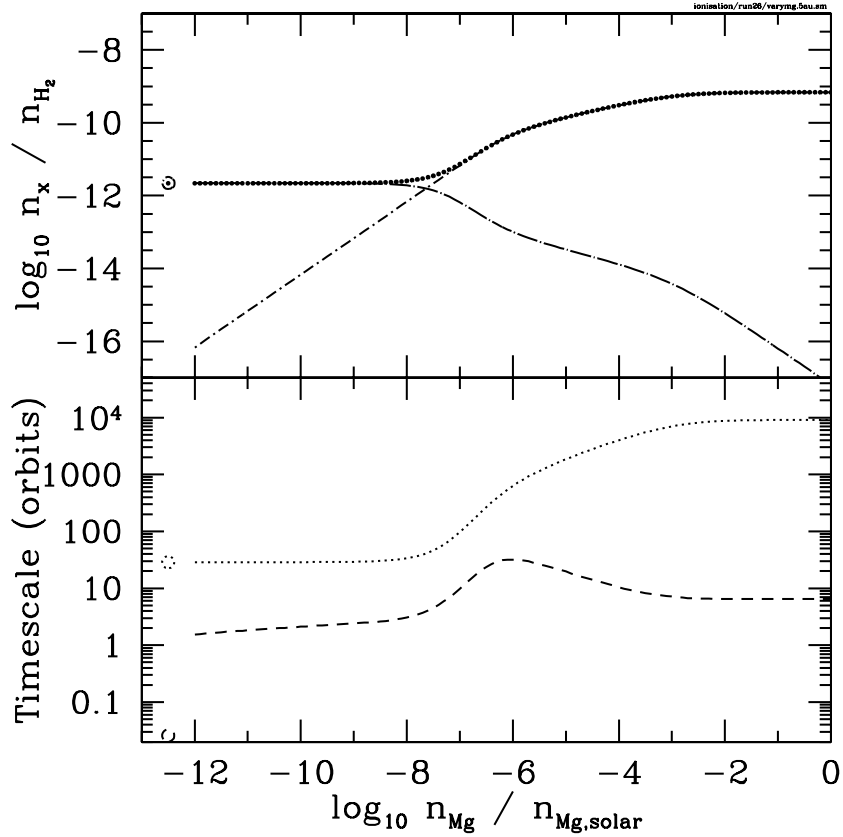}
  \figcaption{Midplane charged particle abundances (top panel) and
    timescales (bottom) versus gas-phase magnesium abundance at 5~AU.
    Symbols are as in figure~\ref{fig:mg1au}: above, H$_2^+$ is shown
    by a long-dash-dotted line, Mg$^+$ by a short-dash-dotted line and
    electrons by small circles.  Below, the recombination timescale is
    shown by a dashed line and the Ohmic dissipation time per scale
    height by a dotted line.  Circles near the left-hand edge mark the
    values when magnesium is absent.
    \label{fig:mg5au}}
\end{figure}

Figure~\ref{fig:z5au} shows the height dependence of the timescales at
5~AU.  The assumed gas-phase magnesium abundance is 1\% of Solar, near
the equilibrium value for thermal adsorption and desorption
(section~\ref{sec:disk}).  MRI turbulence is expected at all heights
because the dissipation is slow.  The relatively high midplane
electron fraction $x_e=6\times 10^{-10}$ is due to faster ionization
than at 1~AU: cosmic rays reaching $z=0$ pass through a column of just
76 g~cm$^{-2}$.  The turbulent mixing is slower than the recombination
in the interior, so the ionization profile will be unaffected by
mixing.  Midplane magnesium abundances greater than $10^{-6}$ Solar
yield dissipation slow enough to permit the MRI to develop
(figure~\ref{fig:mg5au}).
%\clearpage

%\clearpage
%%%%%%%%%%%%%%%%%%%%%%%%%%%%%%%%%%%%%%%%%%%%%%%%%%%%%%%%%%%%%%%%%%%%%%%%%%%%%%%
\section{MAGNETIC ACTIVITY\label{sec:mixing}}

According to the timescale estimates in section~\ref{sec:timescales},
layered accretion is expected at 1~AU, but turbulent mixing could be
fast enough to reduce or dispense with the midplane dead layer.  At
5~AU, the whole thickness of the disk will be active.  Here we explore
the magnetic activity using resistive MHD calculations with and
without time-dependent ionization.

\subsection{Turbulent Mixing at 1~AU\label{sec:mixing1au}}

\begin{figure}
  \epsscale{0.7}
  \plotone{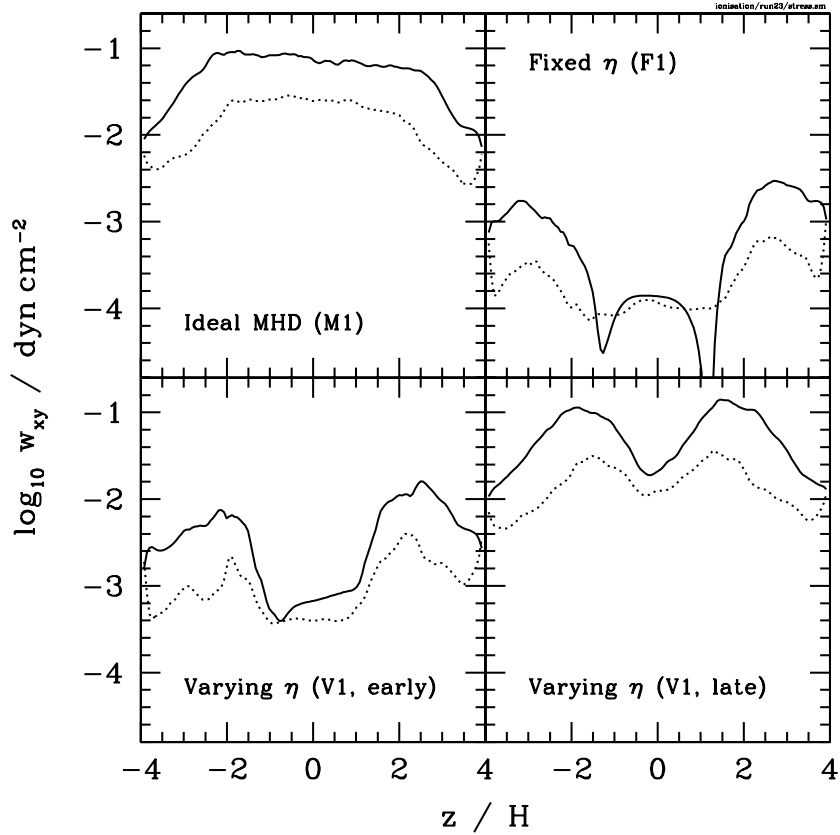}
  \figcaption{Mean accretion stress versus height in MHD calculations
    centered at 1~AU.  The magnetic stress $-B_xB_y/4\pi$ is shown by
    solid lines, the hydrodynamic stress $\rho v_x\delta v_y$ by
    dotted lines; $\delta v_y$ is the azimuthal velocity relative to
    the background shear flow.  The four panels show results from an
    ideal MHD run (top left), a resistive case with a fixed
    resistivity profile $\eta(z)$ (top right), and early and late
    stages from a run with time-varying resistivity (bottom left and
    right, respectively).  The early results are averaged from 10 to
    60~years, the late results from 100 to 150~years.  The gas-phase
    magnesium abundance is Solar.
    \label{fig:stressvsz1au}}
\end{figure}

\begin{figure}
  \epsscale{1.0}
  \plotone{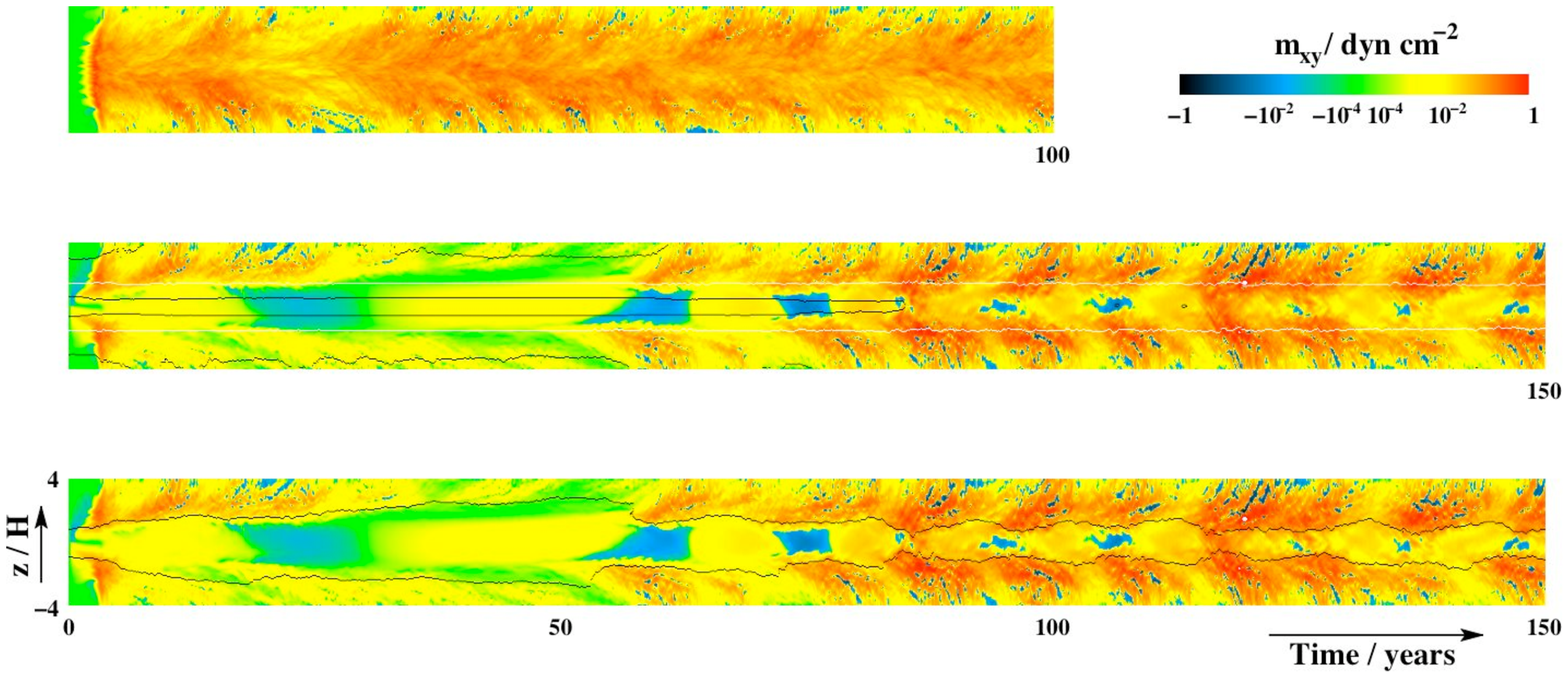}
  \figcaption{Horizontally-averaged magnetic accretion stress versus
    height and time at 1~AU in the ideal MHD calculation M1 (top
    panel) and the resistive MHD version with time-dependent chemistry
    V1 (two lower panels).  The pair of black curves nearest the
    midplane in the middle panel shows where the magnetic Reynolds
    number $Re = c_s^2/(\eta\Omega) = 100$, the white lines $Re=1000$
    and the outer black curves $Re=10^4$.  Black curves in the bottom
    panel mark where the Lundquist number $Lu = v_{Az}^2/(\eta\Omega)$
    is unity.  The common color scale is double logarithmic, with reds
    for positive stresses, blues for negative stresses and green for
    stresses within $10^{-4}$~dyn~cm$^{-2}$ of zero.
    \label{fig:criteria}}
\end{figure}

\begin{figure}[tb!]
  \epsscale{0.7}
  \plotone{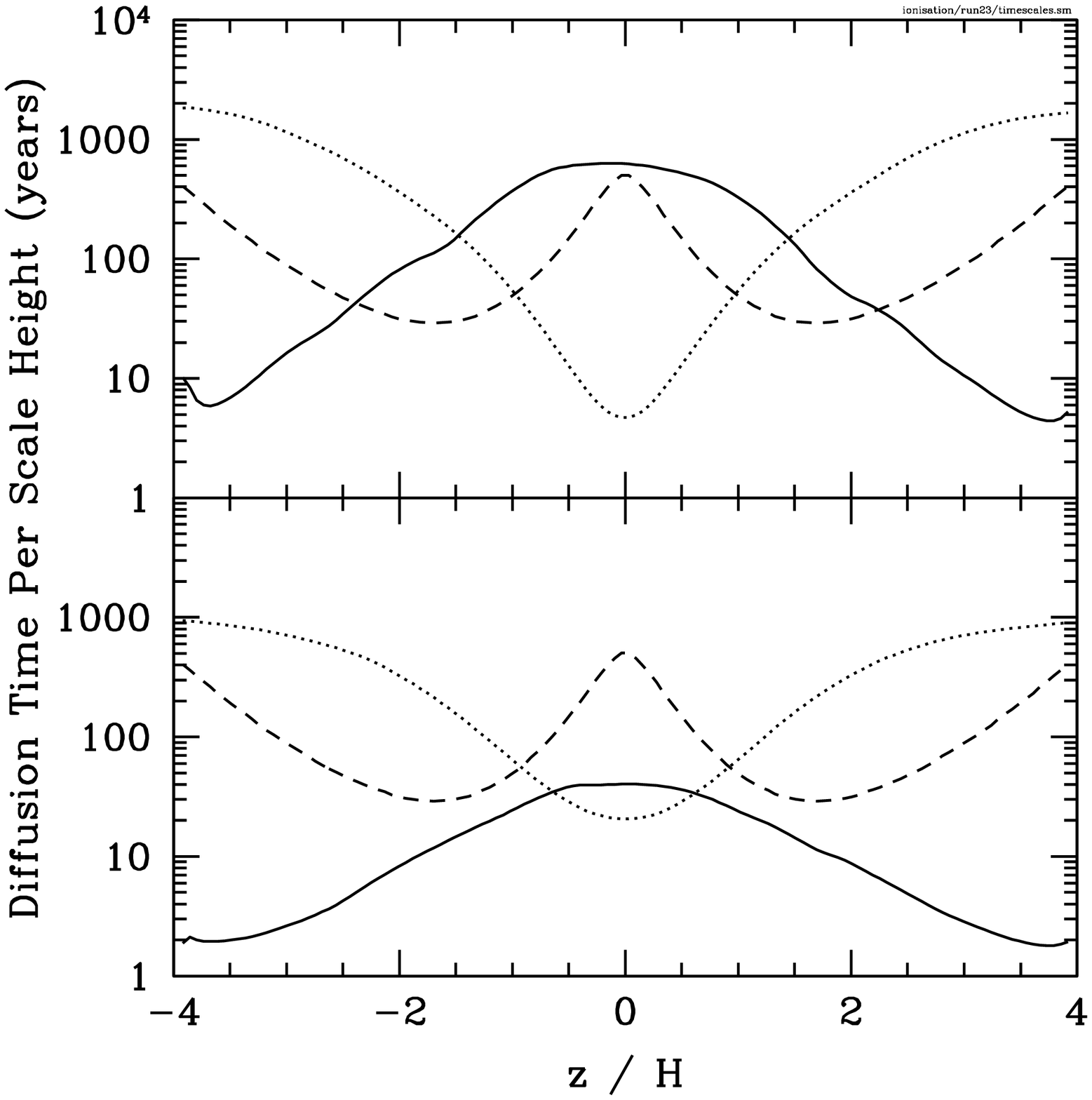}
  \figcaption{Timescales for Ohmic dissipation (dotted), recombination
    (dashed) and turbulent mixing (solid) versus height at 1~AU in the
    MHD calculation V1 with time-dependent ionization.  The results
    are averaged from 10 to 60~years (top panel) and from 100 to
    150~years (bottom panel).  \label{fig:z1auohmic}}
\end{figure}

In calculation F1 the resistivity is a fixed function of height taken
from the equilibrium ionization state in the initial hydrostatic disk
model.  The resulting flow has active surface layers and a midplane
dead zone (figure~\ref{fig:stressvsz1au}).  The stresses in the active
layers are due mostly to magnetic forces, with the contribution from
gas pressure gradients three to six times smaller, while at the
midplane the total stress is an order of magnitude weaker and the
magnetic and hydrodynamic parts are almost equal.  The midplane
magnetic stresses result from net radial and azimuthal fluxes that
build up as fields of the opposing signs escape through the open
vertical boundaries.  Similar calculations by \cite{fs03} showed still
weaker midplane magnetic stresses because the vertical boundaries were
periodic or impermeable and the net radial magnetic flux remained
zero.  Like \cite{fs03} we observe hydrodynamic stresses in the dead
zone due to sound waves propagating away from the active layers.
%\clearpage
%\clearpage
In run V1 the reaction network is solved together with the MHD
equations.  Electrons and ions are carried from place to place in the
flow and the resistivity varies with all three space coordinates and
time.  For the first 60~years the flow again consists of active
surface layers and a midplane dead zone
(figures~\ref{fig:stressvsz1au} and~\ref{fig:criteria}).  The magnetic
fields are weaker overall than the ideal-MHD case M1, leading to
slower mixing.  Recombination is faster than mixing within two scale
heights of the midplane (figure~\ref{fig:z1auohmic}), so the
resistivity approaches its equilibrium profile from
figure~\ref{fig:z1au} even when the time-dependent chemistry is
included.  The dissipation time differs most from
figure~\ref{fig:z1au} in the active layers, where mixing makes the
electron fraction and resistivity more nearly uniform.  The turbulent
mixing has little effect on the size of the midplane dead zone.
%\clearpage

\begin{figure}
  \epsscale{0.7}
  \plotone{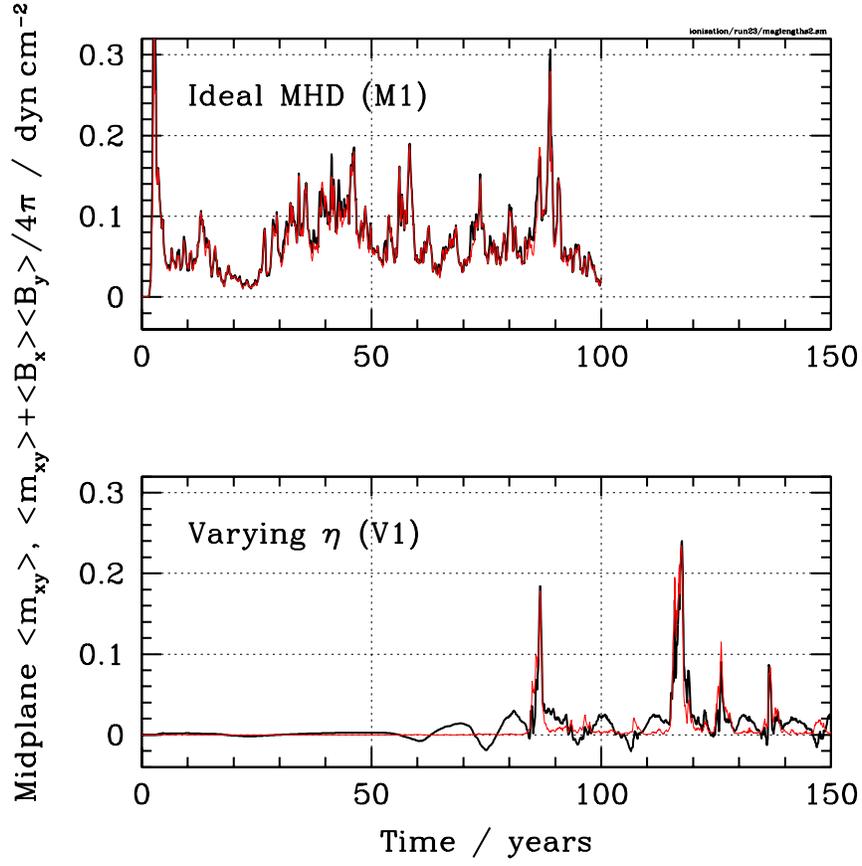}
  \figcaption{Midplane magnetic accretion stresses at 1~AU in the
    ideal MHD calculation M1 (top panel) and the resistive MHD
    calculation V1 (bottom panel).  Stresses due to the entire
    magnetic field are shown by thick curves, those due to small-scale
    fields alone by thin red curves.  \label{fig:maglengths}}
\end{figure}

\begin{figure}
  \epsscale{0.7}
  \plotone{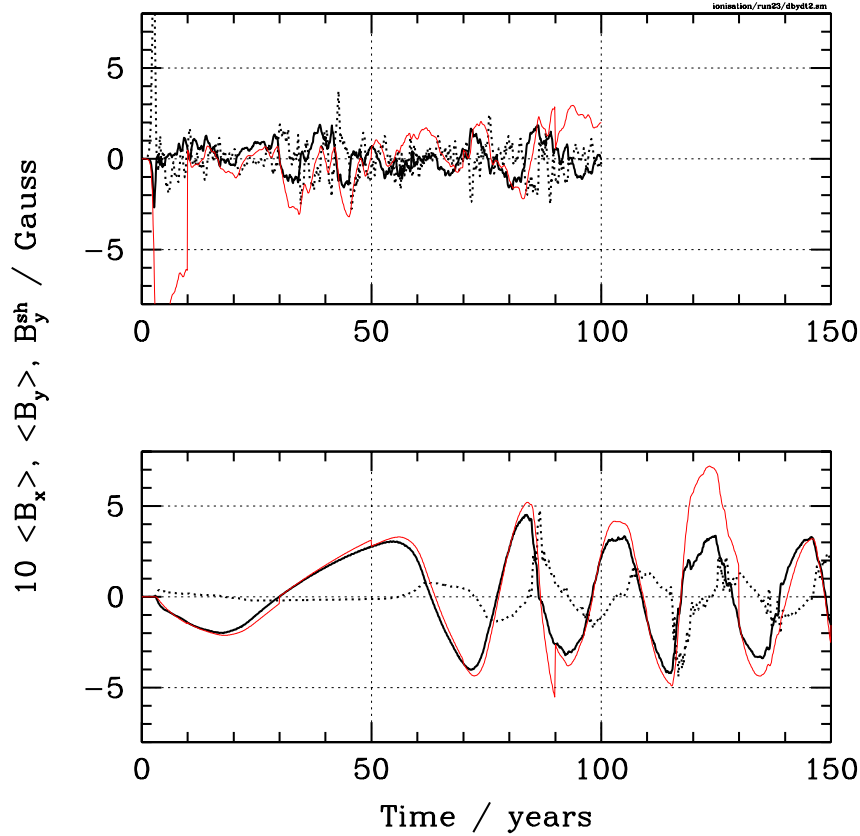}
  \figcaption{Midplane magnetic fields versus time at 1~AU in the
    ideal MHD calculation M1 (top panel) and the resistive MHD
    calculation V1 (bottom panel).  The azimuthal component is shown
    by thick solid curves, the radial component by dotted curves.
    Both are horizontally-averaged and the radial component is
    multiplied by ten so its variations can be read on the same scale.
    The azimuthal field that would result from shear acting on the
    radial field (eq.~\ref{eqn:shear}) is indicated by thin red
    curves.  \label{fig:dbydt2}}
\end{figure}

%\clearpage
The situation changes around 60~years, when a period of stronger
turbulence leads to faster mixing.  Ionization at the midplane
increases and the active regions expand toward $z=0$.  The midplane
electron fraction peaks at 87~years at $x_e = 1.2\times 10^{-11}$ and
then varies around $10^{-11}$ till the run ends at 150~years.  The
flow settles into a new state in which the mixing at the midplane is
faster than the recombination, but still slower than the dissipation
of magnetic fields (figure~\ref{fig:z1auohmic}).  The fast-growing MRI
modes are adequately spatially resolved throughout the volume.
Outside two scale heights the characteristic wavelength averages ten
or more grid zones in the vertical direction due to the large Alfven
speeds, while nearer the midplane the linear modes not damped by the
resistivity have vertical wavelengths six zones or greater.  The mean
accretion stresses are similar to the ideal-MHD calculation M1 in the
surface layers outside $|z|=H$, but a few times weaker at the midplane
(figure~\ref{fig:stressvsz1au}).  The midplane is magnetized and has a
significant total accretion stress that is 0.2\% of the local gas
pressure.  However the magnetic activity differs in character from the
surface layers and is also unlike the midplane in the ideal-MHD run M1
in several respects:
\begin{enumerate}
\item The magnetic field shows few features smaller than a quarter of
  a scale height, as expected from the dissipation timescale of
  20~years per scale height.
\item The magnetic stress $m_{xy}=-B_xB_y/4\pi$ is due mostly to
  large-scale fields, similar to the dead zone during the early part
  of the calculation, rather than small-scale fields as in the
  ideal-MHD run (figure~\ref{fig:maglengths}).  The
  horizontally-averaged magnetic stress $\left<m_{xy}\right>$ roughly
  equals the stress from the horizontally-averaged fields,
  $-\left<B_x\right>\left<B_y\right>/4\pi$, except for two periods of
  a few orbits around 87 and 117~years.
\item The horizontally-averaged stress varies around zero, while in
  the ideal-MHD case it is strictly positive
  (figure~\ref{fig:maglengths}).
\item The mean ratio of magnetic to hydrodynamic stresses is just 1.9,
  intermediate between the ratios in the dead zone and active layers
  during the early part of the calculation.
\item The mean midplane pressure in the vertical component of the
  magnetic field is 125~times smaller than that in the azimuthal
  component, contrasting with the ratios around 30 found in run M1 and
  also typical of past MRI turbulence calculations \citep{ms00}.  The
  relative strength of the azimuthal field suggests a generation
  mechanism distinct from the MRI.
\item The mean azimuthal magnetic field evolves mostly through shear
  acting on the mean radial field.  If shear were the only effect, the
  azimuthal field at a time $t>t_0$ would be
  \begin{equation}\label{eqn:shear}
  B_y^{sh}(t) = B_y(t_0) - \int_{t_0}^t \frac{3}{2} \Omega B_x\,dt.
  \end{equation}
  The mean azimuthal field is plotted together with $B_y^{sh}$ in
  figure~\ref{fig:dbydt2}.  The departure from the shear curve is
  shown over 20-year intervals by performing the integration from
  $t_0=0$ up to 10~years, from $t_0=10$ up to 30~years, from $t_0=30$
  up to 50~years and so on.  The field closely tracks the variation
  expected from shear throughout run V1, while the two curves are more
  weakly correlated in the ideal-MHD version M1.
\end{enumerate}
These results indicate the magnetic fields evolve as follows.  The
fields delivered to the interior suffer Ohmic dissipation of
small-scale structures, leaving a weak, almost uniform net radial
component.  Although the MRI is ineffective near the midplane due to
the large resistivity, the shear produces a net azimuthal field,
leading to an accretion stress that can be positive or negative.
However the stress is negative for only short periods and when
time-averaged is positive, driving accretion
(figure~\ref{fig:stressvsz1au}).  The shear-generated fields are
expelled both up and down, leading to correlated activity in the two
surface layers unlike run M1 (figure~\ref{fig:criteria}).  The
stresses and the azimuthal field strength peak about every ten orbits.
Azimuthal field variations with similar periods occurred in MHD
calculations by \cite{bd97}.  The variations depend little on the
domain height provided the box extends to levels where the magnetic
pressure exceeds the gas pressure, according to ideal MHD tests
\citep{tw06}.  Occasionally when the midplane azimuthal field is
strongest it erupts, leading to the production of a large midplane
radial field as at 87 and 117~years, and then is destroyed within a
few orbits (figure~\ref{fig:dbydt2}).  The magnetic energy dissipated
in these outbursts per unit area per unit time is comparable to the
radiative flux from the surface of the viscous model disk with the
same stress and mass column, suggesting such events could be directly
observed.

%\clearpage
\subsection{Turbulence at 5~AU\label{sec:mixing5au}}

\begin{figure}
  \epsscale{0.7}
  \plotone{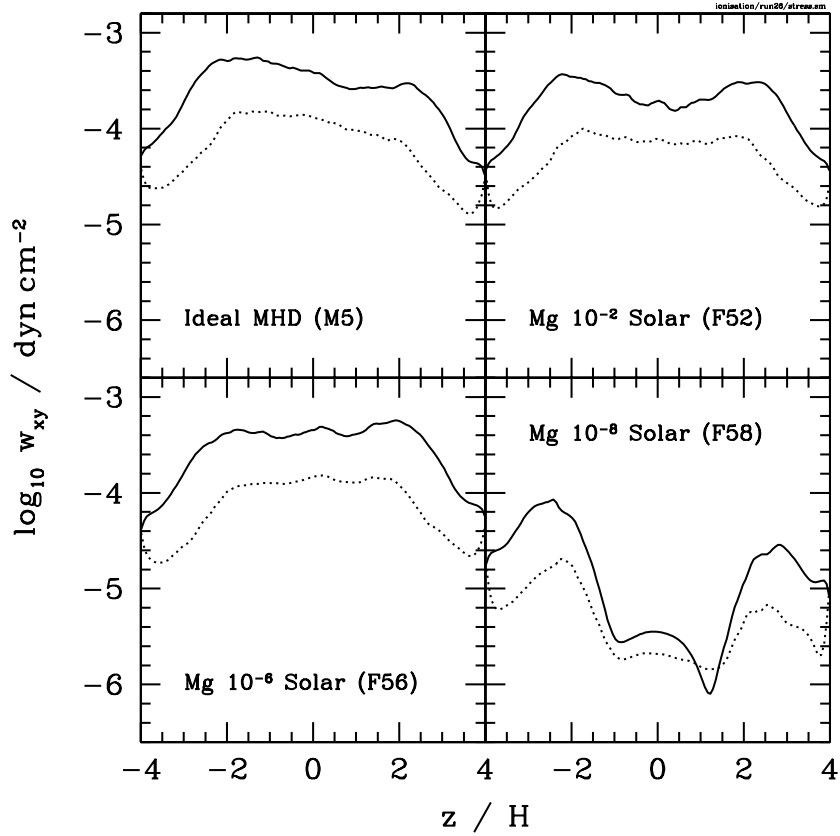}
  \figcaption{Mean accretion stress versus height in four MHD
    calculations centered at 5~AU.  The magnetic stress is shown by
    solid lines, the hydrodynamic stress by dotted lines.  The four
    panels show results from an ideal MHD calculation (top left) and
    calculations with a fixed resistivity profile and gas-phase
    magnesium abundances 0.01 (top right), $10^{-6}$ (bottom left) and
    $10^{-8}$ times Solar (bottom right).  The results are averaged
    from 10 to 100~orbits.
    \label{fig:stressvsz5au}}
\end{figure}

Turbulence fills the entire disk thickness in the shearing-box
calculations at 5~AU that have gas-phase magnesium abundances
$10^{-6}$ Solar and greater.  No dead zone is present and the
accretion stresses have almost identical mean vertical profiles in the
cases M5, F52 and F56 (figure~\ref{fig:stressvsz5au}).  By contrast,
the run F58 with magnesium abundance $10^{-8}$ Solar has a dead zone
extending two scale heights from the midplane.  The midplane magnetic
dissipation time per scale height in F58 is just 34~orbits, compared
with 630~orbits in run F56 (figure~\ref{fig:mg5au}).  Recombination at
the midplane takes only three orbits and is much faster than even the
mixing in the ideal MHD run (figure~\ref{fig:z5au}), so turbulent
mixing will not expunge the dead zone in this case.

%\clearpage
%%%%%%%%%%%%%%%%%%%%%%%%%%%%%%%%%%%%%%%%%%%%%%%%%%%%%%%%%%%%%%%%%%%%%%%%%%%%%%%
\section{ACTIVITY CRITERIA\label{sec:criteria}}

Simple criteria for the presence of turbulence driven by the MRI are
extremely useful in developing protostellar disk models without the
investment of time needed for three-dimensional MHD calculations.  Two
such criteria are tested in this section.

An Ohmic resistivity quenches the linear growth of the MRI if the
magnetic fields diffuse across the characteristic wavelength
$\lambda_c$ faster than the perturbations grow.  The diffusion time is
the wavelength squared divided by the resistivity, yielding a
criterion for instability on vertical fields
\begin{equation}
Lu \equiv {v_{Az}^2 \over \eta\Omega} > 1 \label{eqn:lundquist}
\end{equation}
in agreement with linear analyses \citep{j96,sm99}.  The left-hand
side of eq.~\ref{eqn:lundquist} is the dimensionless Lundquist number.
In the fully-developed turbulence, the growth is most easily halted by
diffusion perpendicular to the midplane because on average the field
reverses direction most often per unit length along the $z$-direction.
Numerical unstratified shearing-box results indicate that
eq.~\ref{eqn:lundquist} applies to the turbulence also
\citep{si01,ss02b}.

The second activity criterion is constructed using the diffusion time
across the density scale height $H$.  The resulting magnetic Reynolds
number $Re \equiv c_s^2 / (\eta\Omega)$ is independent of the field
strength and so the value required for instability depends on the
fields present.  Turbulence occurred only at Reynolds numbers greater
than $10^4$ with zero net magnetic flux and at Reynolds numbers as low
as 100 with a net vertical magnetic flux, in unstratified shearing-box
calculations by \cite{fs00}.

The edges of the dead zone according to the two activity criteria are
shown in figure~\ref{fig:criteria}.  In run V1 with the reaction
network, the Reynolds number is less than $10^4$ throughout the domain
after about 60~years and greater than 100 everywhere from 85~years on.
By the zero net flux criterion $Re>10^4$ the dead zone fills the
domain, which is clearly not the case.  On the other hand by the
vertical field criterion $Re>100$, the dead zone is completely absent
and MRI activity extends to the midplane, contradicting the evidence
outlined in section~\ref{sec:mixing1au}.  It may be that neither of
these criteria applies to our calculations, which have zero net
vertical but non-zero net radial and azimuthal magnetic fluxes.  An
intermediate critical Reynolds number of 1000, corresponding to a
dissipation time of 160~orbits per scale height, more nearly
approximates the boundaries of the MRI active regions, but fails to
respond to the changes in magnetic field strength over the course of
the calculation.  In contrast to these three possible critical
Reynolds numbers, as shown in the bottom panel of
figure~\ref{fig:criteria} a Lundquist number unity accurately traces
the edges of the MRI turbulent layers at all times in run V1.

The activity criteria are similar at 5~AU.  The mean midplane Reynolds
number is $6\times 10^4$ in run F52, 4000 in run F56 and 200 in run
F58.  Only the last of these has a dead zone
(figure~\ref{fig:stressvsz5au}), consistent with a critical Reynolds
number of 1000 for this situation.  The Lundquist number is greater
than unity at all heights in the first two cases and greater than
unity in run F58 only outside the MRI dead zone, again obeying
eq.~\ref{eqn:lundquist}.

Overall, the Reynolds number suffers from the difficulty that its
critical value must be determined for each magnetic field
configuration using 3-D MHD calculations, while the Lundquist number
always has a critical value unity.  The Lundquist criterion is
preferable because it explicitly shows the magnetic field dependence.
Neither dimensionless number bypasses the basic uncertainty regarding
the net magnetic flux, which is an input parameter for the shearing
box calculations.  However once a disk model and field are chosen, the
Lundquist criterion eq.~\ref{eqn:lundquist} immediately gives the size
of the dead zone.

%%%%%%%%%%%%%%%%%%%%%%%%%%%%%%%%%%%%%%%%%%%%%%%%%%%%%%%%%%%%%%%%%%%%%%%%%%%%%%%
\section{CONCLUSIONS\label{sec:conc}}

We investigated whether the dead zone is eliminated from the
minimum-mass Solar nebula under favorable conditions.  Cosmic rays are
assumed to reach the disk surface, recombination on grains is absent
and metals are found in the gas phase.  The most restrictive of these
assumptions is the lack of small grains.  Particles 0.1~$\mu$m in
radius can significantly affect the thickness of the dead zone at 1~AU
even with dust-to-gas mass ratios 100~times lower than the
interstellar value \citep{sm00}.  The gas-phase metal abundance is a
less important parameter and can be several orders of magnitude below
its nominal value with little change in the dead zone size.

Timescale estimates and numerical resistive MHD calculations both show
that the whole of the disk thickness at 5~AU is active if the
gas-phase magnesium abundance is greater than $10^{-6}$ times Solar,
consistent with results at a similar column density from \cite{in06a}.
The problem is more severe at 1~AU, where few ionizing particles
penetrate to the midplane if the column density is 1700~g~cm$^{-2}$.
In chemical equilibrium the dead zone thickness declines with
increasing gas-phase magnesium abundance but asymptotes at values
above $10^{-4}$ times Solar with most of the mass column still
magnetically inactive.  However ionized gas can reach the interior
through turbulent mixing, because the recombination and mixing
timescales are similar.  In an MHD calculation including
time-dependent ionization chemistry, free electrons arrive at the
midplane fast enough to maintain a weak coupling to the magnetic
fields.  The midplane remains linearly stable against the MRI because
of the large resistivity, but receives magnetic fields generated in
the surface layers.  The midplane fields are smooth while those in the
surface layers are tangled.  The boundary between the two lies at the
height where the MRI is marginally stable and the Lundquist number is
unity.  Radial fields of one sign preferentially escape from the disk
surfaces, leaving the midplane with a net radial magnetic flux of the
opposite sign and leading to the generation of azimuthal fields
through shear.  The resulting accretion stress varies around zero, but
on average is positive and just a few times less than the stress in
the active layers.

These results demonstrate that the dead zone can be eradicated and the
midplane gas flow toward the star even if the MRI occurs only in the
surface layers.  Magnetic forces can provide angular momentum transfer
throughout the planet formation region once small grains have been
removed.  However the midplane flow at 1~AU is fundamentally different
from the usual picture.  It is magnetorotationally stable and has
magnetic stresses that are uniform on scales up to the disk thickness.
The presence of internal stresses yields a greater overall accretion
rate than the basic layered model, permitting a stronger dependence on
the stellar mass as required by brown dwarf and T~Tauri star
measurements \citep{hd06}.  Time variability in our calculations is
due to fluctuations in the magnetic field strength.  Longer-period
accretion variability can arise from the pileup of material reaching
the dead zone, even in the presence of midplane stresses \citep{wg06};
these effects are not included in our calculations.  The smooth
internal magnetic fields have pressures up to 5\% of the gas pressure
and are strong enough to modify the orbital migration of Earth-mass
protoplanets through their effects on the corotation resonance
\citep{t03}.  Other potentially important effects in the dead zone
are the transport of magnetic flux over large radial distances and the
interactions between disk annuli carrying opposing magnetic fields.
These can be explored with future global MHD calculations.

%%%%%%%%%%%%%%%%%%%%%%%%%%%%%%%%%%%%%%%%%%%%%%%%%%%%%%%%%%%%%%%%%%%%%%%%%%%%%%%
\acknowledgments

Portions of this work were carried out at the Jet Propulsion
Laboratory, California Institute of Technology using the JPL
Supercomputing and Visualization Facility.  Support was provided by
the JPL Research and Technology Development and NASA Solar System
Origins Programs.

%%%%%%%%%%%%%%%%%%%%%%%%%%%%%%%%%%%%%%%%%%%%%%%%%%%%%%%%%%%%%%%%%%%%%%%%%%%%%%%

\end{document}